\documentclass[%
 reprint,
groupedaddress,
 amsmath,amssymb,
 aps, prl,
]{revtex4-2}

\usepackage{graphicx}
\usepackage{dcolumn}
\usepackage{bm}
\usepackage{hyperref}
\usepackage[mathlines]{lineno}
\usepackage{romannum}

\begin{document}


\title{Internal Structure of Metal Vacancies in Cubic Carbides}
%
\author{Ekaterina Smirnova}
\email{ekaterina.smirnova@misis.ru}
\affiliation{
National University of Science and Technology MISIS, 119049 Moscow, Russia
}
\author{Mehdi Nourazar}
\email{mehdin@kth.se}
\affiliation{
KTH Royal Institute of Technology, SE-100 44 Stockholm, Sweden
}%
\author{Pavel A. Korzhavyi}
\email{pavelk@kth.se}
\affiliation{
KTH Royal Institute of Technology, SE-100 44 Stockholm, Sweden
}

\date{\today}

\begin{abstract}
A combinatorial approach is employed to investigate the atomic and electronic structures of a metal vacancy in titanium carbide. It turns out that the usual relaxed geometry of the vacancy is just a metastable state representing a local energy minimum. Using \textit{ab initio} calculations and by systematically searching through the configurational space of a Ti monovacancy, we identify a multitude of local minima with reconstructed geometry that are lower in energy. Among them, there is a planar configuration with two displaced carbons forming a dimer inside the vacancy. This structure has the optimal number and order of C--C bonds making it the global minimum. Further calculations show that this reconstructed geometry is also the ground state of metal vacancies in other carbides such as ZrC, HfC, and VC. The reconstructed metal vacancies are characterized by localized electron states due to the relatively short C--C bonds. The defect states lie just below the upper and lower valence bands. The existence of reconstructed vacancy configurations is essential for understanding the mechanism of metal self-diffusion in transition-metal carbides.  
\end{abstract}

\maketitle


Carbides and nitrides of group 4 and 5 transition metal (TM) elements crystallize in the cubic B1 (Halite) structure typical of ionic crystals such as NaCl. Strong ionic-covalent bonding among the metal and non-metal atoms gives these materials their high melting points, great hardness, and high chemical stability. At the same time, metallic bonding in the TM carbides and nitrides makes them electrically conductive and enhances their thermal conductivity \cite{Toth71,Shabalin2019,Shabalin2020}. Thanks to this combination of ceramic and metallic properties, cubic carbides find numerous applications as materials for coatings and cutting-tools \cite{Holleck86}, as well as in nuclear and aerospace industries \cite{WsWill97,Fahrenholtz2014}. 

Diffusion mechanisms in TM carbides are of fundamental importance in understanding their functional and mechanical properties at elevated temperatures \cite{KKebler64,ISpivak74,JChermant80,CJSmith18}. For instance, metal (Me) atom diffusion in TM carbides is of direct relevance to the phase separation in mixed carbides such as (Ti,Zr)C that hardens the material by producing an ultrafine mixture of TiC- and ZrC-enriched regions  \cite{OKnotek89,Ma2016,Yildiz2022}. Diffusion in TM carbides has been thoroughly studied experimentally in connection with the problem of high-temperature creep (time-dependent deformation under applied load) of these materials \cite{Ssarian68,Ssarian69,VNZagryazkin69,DKohlstedt70,RAAndriev71,FJJVLoo89,Andrievskii2011}. 

The commonly accepted view on diffusion in Halite-structured compounds is that they have a Schottky defect structure where the thermal as well as the constitutional (structural) point defects are vacancies. Indeed, non-metal vacancies are abundant in TM carbides and nitrides where they cause large deviations from stoichiometry \cite{Hultman2000,Gusev2001,Andersson08}. Certain TM oxides are found to contain structural vacancies on both, metal and non-metal, sublattices \cite{Watanabe67,Valeeva2000,Valeeva2001}. 

\begin{figure}
\includegraphics[width=8cm]{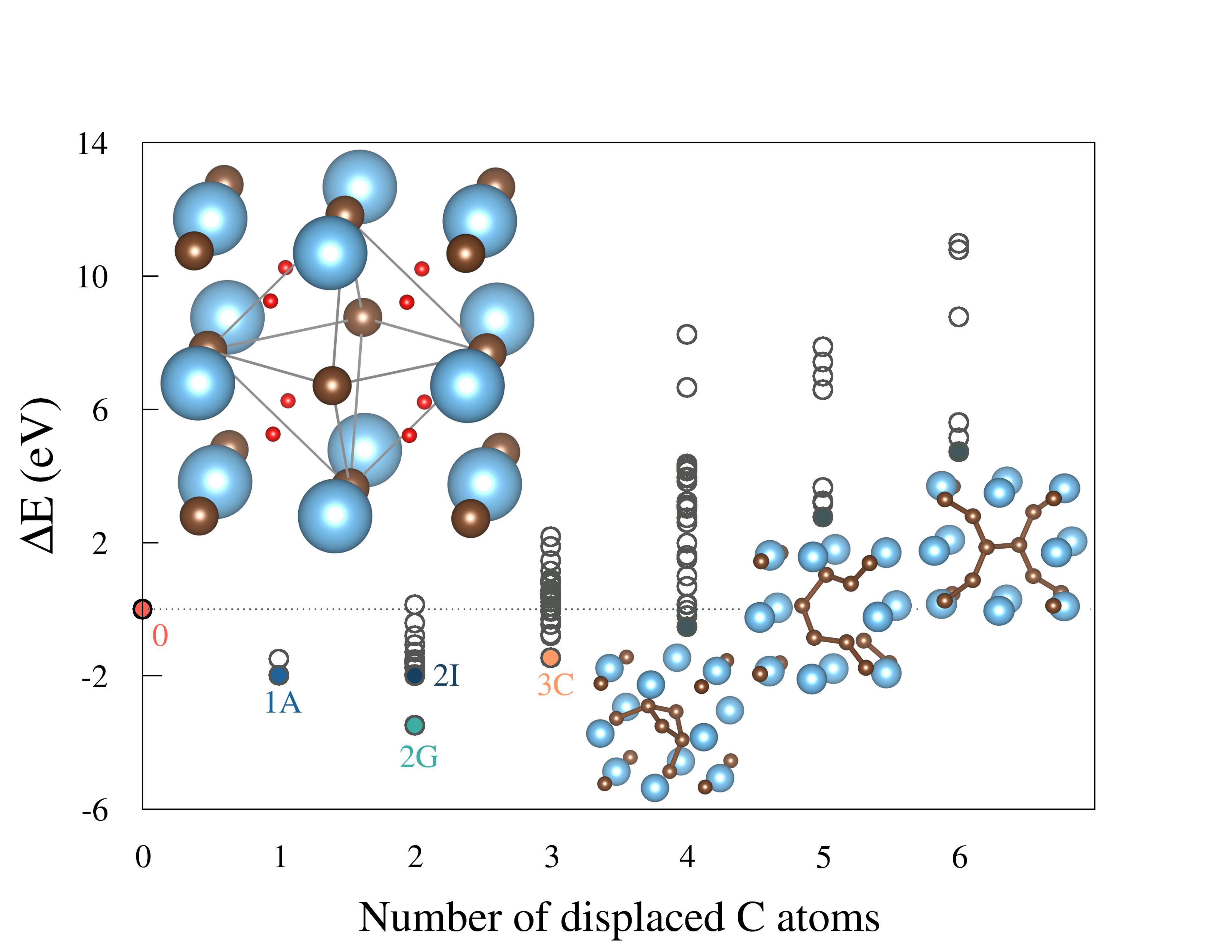}%
\caption{(a) Reconstruction energy $\Delta E$ of a metal vacancy in TiC as a function of the number of displaced C atoms. Filled symbols with labels denote low-energy structures to be analyzed in the text. Unlabelled filled symbols correspond to structures shown below. Upper left corner: Unreconstructed structure 0 of a metal vacancy in TiC: Ti atoms (blue spheres), an octahedron of six nearest-neighbor C atoms (brown spheres connected by lines), and eight tetrahedral interstitial sites (red spheres). All structures are visualized using VESTA \cite{Momma2011}   \label{fig1}}
\end{figure}

While Schottky defect structure in TM nitrides and oxides is basically confirmed by \textit{ab initio} calculations \cite{Andersson05,Tsetseris2010,Razumov15,Gambino2017}, the situation in carbides is controversial \cite{Ltsets08,HmPinto09,Razumov11,Xiao20,Rofiq21}. On the one hand, the stability of carbon vacancies and related thermodynamic properties of sub-stoichiometric carbides such as TiC$_{1-x}$ are correctly reproduced. On the other hand, \textit{ab initio} calculated formation energy of a metal vacancy (Fig. \ref{fig1}) \textit{via} a Schottky defect in TiC is almost 8 eV, \cite{Razumov11,Razumov13,Razumov15,Xiao20}, far too high if one remembers that the experimental value for activation energy of metal self-diffusion in Ti and Zr carbides, about 7.5 eV  \cite{Ssarian69,Andrievskii2011}, must also include a vacancy migration energy of at least 3.5 eV \cite{Ltsets08,Razumov11,Xiao20}. 

Several attempts have been made to understand how a metal vacancy can lower its formation energy. First, Razumovskiy \textit{et al}. \citep{Razumov13} noticed that carbon vacancies Va$_{\rm C}$, which are abundant in sub-stoichiometric TiC$_{1-x}$, exhibit a strong affinity to a metal vacancy Va$_{\rm Me}$ and tend to fully surround it to form a 6Va$_{\rm C}$--Va$_{\rm Me}$ vacancy cluster, thereby lowering the formation energy of a metal vacancy down to about 3 eV. Such a vacancy cluster can migrate in TiC \textit{via} a series of correlated atom--vacancy jumps, with a migration energy of some 3.5 to 4 eV. An alternative diffusion mechanism, mediated by an interstitial--vacancy cluster (a Ti dumbbell terminated by two C vacancies) was proposed by Sun \textit{et al}. \citep{Wsun15,Wsun19}. Although the two mentioned defect cluster mechanisms may account for the observed activation energy of metal self-diffusion in carbides, they fall short in explaining the experimentally reported pre-factor for metal self-diffusion coefficient. The pre-factor is anomalously high, suggesting a high activation \textit{entropy} of about 10 to 14.5 $k_{\rm B}$, \citep{Ssarian69,RAAndriev71} which is difficult to reconcile with the highly correlated processes of cluster migration \citep{Razumov13,Wsun19,Rofiq21}.

These difficulties of atomistic modeling of metal diffusion in group 4 TM carbides look especially surprising since straightforward \textit{ab initio} evaluations of the monovacancy mechanism of metal self-diffusion in nitrides of the same elements have been successful \cite{Tsetseris2010,Razumov15,Gambino2017}. 
Furthermore, \textit{ab initio} calculations \cite{Andersson05} for vacancy defects in titanium monoxide TiO show, in agreement with experiment \cite{Watanabe67,Valeeva2000,Valeeva2001}, that both metal and non-metal vacancies are stable (structural) defects, suggesting a negative formation energy of a Schottky defect in perfect B1-TiO. Indeed, the crystal structure of TiO may be viewed as an incompletely filled NaCl-type lattice where about 1/6 of the metal and non-metal sites are vacant. The vacancies are disordered in the high-temperature $\beta$-phase but form an ordered motif in the low-temperature $\alpha$-TiO phase \cite{Valeeva2001}.

To demonstrate the chemical trend, we re-evaluated the Schottky defect formation energy in otherwise defect-free NaCl-type TiC, TiN, and TiO compounds. The calculations were done using the Vienna ab-initio simulation package (VASP) \citep{Kresse96}, projector augmented wave (PAW) type pseudopotentials \cite{Blochl94,Gkresse99}, the Perdew-Burke-Ernzerhof form of exchange-correlation functional \citep{Perdew96}, and a plane-wave basis set with a cutoff energy of 500 eV. Brillouin zone integration was carried out on a 5$\times$5$\times$5 Monkhorst-Pack mesh of special $k$-points using the Fermi smearing technique with an electronic temperature of 1000~K. The 216-site supercells with and without a dissociated vacancy pair were fully relaxed using convergence criteria 10$-6$ eV for the energy and 10$-4$ eV/\AA for the forces. In agreement with previous calculations \cite{Razumov11,Razumov15,Andersson05}, the formation energy of a Schottky defect is found to regularly decrease from a high positive value of 7.42 eV for TiC, through a moderate value of 2.14 eV for TiN, to a negative value of $-4.48$ eV for TiO. This chemical trend agrees with the positron annihilation measurements detecting  only non-metal vacancies in TiC \cite{Rempel1998}, but both metal and non-metal vacancies in TiO \cite{Valeeva2007}.

Taking into account that TiC and TiO are mutually miscible, and that TM carbides are often synthesized from the oxide precursors, oxygen on the carbon sublattice O$_{\rm C}$ is expected to be a common impurity in TiC. While checking whether a Ti vacancy Va$_{\rm Ti}$ near an O$_{\rm C}$ impurity lowers its formation energy, and whether a Va$_{\rm C}$--Va$_{\rm Ti}$--O$_{\rm C}$ defect cluster could migrate in O-contaminated TiC crystal, we observed the following phenomenon: a Frenkel C pair (Va$_{\rm C}$--C$_{\rm t}$) consisting of a carbon vacancy and a C atom in the tetrahedral interstitial position (denoted by subscript t) formed spontaneously near the titanium vacancy. This was unexpected because the formation energy of a carbon Frenkel pair in defect-free TiC is positive and large, up to 4 eV \citep{Razumov11,Wsun15}. No spontaneous Frenkel pair formation was ever observed in our previous studies of self-diffusion in TiC involving vacancies, interstitials, and their clusters \cite{Razumov13,Wsun15,Wsun19}. 

To thoroughly check whether Frenkel C pairs can form spontaneously near a Ti vacancy in \textit{pure} TiC, we set up the following enumeration procedure:

\begin{enumerate}\addtocounter{enumi}{-1}
\item A supercell containing 216 sites, with 107 Ti atoms, 108 C atoms, and 1 Ti vacancy, is created and fully relaxed.
\item Some number (ranging from 0 to 6) of the C atoms nearest to the vacancy are displaced into some of the 8 tetrahedral interstitial positions around the vacancy, see Fig. \ref{fig1}, to create Frenkel C pairs.
\item The created atomic structure is internally relaxed using forces to a local energy minimum. 
\item The resulting atomic and electronic structures are recorded and analyzed.
\end{enumerate}
Steps 1 to 3 are repeated to enumerate the 96 geometrically inequivalent possibilities (out of 3003 possible combinations), see Table \ref{tbl1} for details. We note that the enumerated structures with C atoms displaced into the interstitial sites are just starting configurations. In most cases, the atoms simply relaxed towards the nearest local minimum during the lattice relaxation, but some of the initial configurations reconstructed quite severely. Several different starting configurations would reconstruct into the same final structure (sometimes with a different number of displaced C atoms than initially). 

\begin{figure}
\includegraphics[width=9cm]{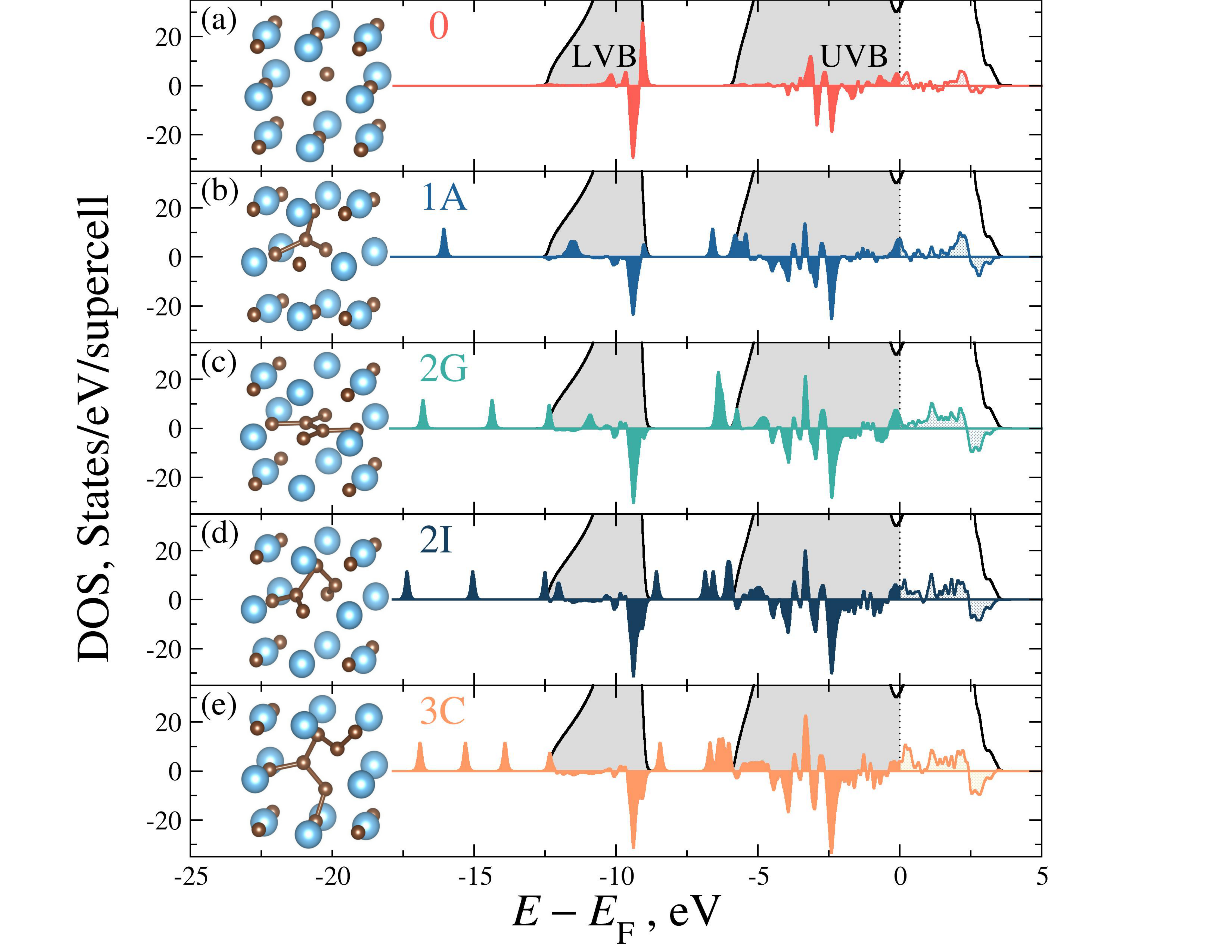}%
\caption{Total (black line) and differential (colored lines) DOS for the unreconstructed (a) and selected reconstructed (b-e) configurations of a metal vacancy in TiC. Shaded areas denote states below the Fermi energy, $E_F$. Insets show the local atomic structures around the vacancy, notations for atomic species are the same as in Fig. \ref{fig1}.  \label{fig2}}
\end{figure}

In these calculations, we employed the same 216-site supercells, a slightly reduced cutoff energy of 400 eV, the same 5$\times$5$\times$5 mesh of $k$-points, and Fermi smearing at an electronic temperature of 2000~K. The final energy and density of states (DOS) calculations for the relaxed structures were done on a denser $k$-mesh of 13$\times$13$\times$13 and at a lower electronic temperature of 500~K. The main results of this systematic search for possible atomic geometries of a Ti vacancy in TiC are presented in Figs. \ref{fig1} and \ref{fig2} and briefly summarized in Table \ref{tbl1}. The analysis below is given for TiC and then extended to metal vacancies in other TM carbides of group 4 and 5. 

\begin{table}[b]%
\caption{\label{tbl1}%
Parameters and main results of ground-state search: $p$ - number of Frenkel pairs (displaced C atoms) into a Ti vacancy in B1-TiC, $N_{in}(p)=C(6,p)C(8,p)$ - number of all possible initial configurations, $n_{in}(p)$ - symmetry-reduced number of initial configurations, $n_{out}(p)$ - number of different structures with $p$ displaced C atoms after the relaxation, and $\Delta E_{min}=\min \left\lbrace E_{out}(p)-E_{out}(0) \right\rbrace$ - the minimum energy of reconstruction, eV.}
\begin{ruledtabular}
\begin{tabular}{lccccccc}
$p:$& 0 & 1 & 2 & 3 & 4 & 5 & 6 \\
\colrule
$N_{in}(p)$ & 1 & 48 & 420 & 1120 & 1050 & 336 & 28 \\
$n_{in}(p)$ & 1 & 2 & 16 & 30 & 34 & 10 & 3 \\
$n_{out}(p)$ & 1 & 2 & 15 & 25 & 32 & 9 & 6 \\
$\Delta E_{min}$ & 0.0 & -1.9 & -3.5 & -1.5 & -0.5 & +3.5 & +4.5 \\
\end{tabular}
\end{ruledtabular}
\end{table}

\paragraph*{Structure 0} We begin our analysis with the usual (unreconstructed) geometry of a Ti vacancy, denoted as '0' to indicate that no C atoms have been displaced into the tetrahedral interstitial positions. This atomic geometry, obtained after the internal relaxation, is depicted in Fig. \ref{fig2}(a), where we also plot the so-called differential density of states (dDOS), introduced in Ref. \cite{Wsun15} to visualize the states created (positive peaks) or removed (negative peaks) from the total density of states as a result of the defect formation. The Figure shows that the C atoms surrounding the vacancy relax slightly away from the vacancy center, and that the Ti vacancy does not produce any defect states outside the the lower valence band (LVB) or upper valence band (UVB) of TiC. Structure 0 turns out to be just one of many metastable states of a Ti vacancy, corresponding to a local minimum of total energy as a function of atomic coordinates. The total energy of structure 0 is chosen here to be zero; the energies of all other studied structures will be expressed relative to it.


\paragraph*{Structure 1A} Two symmetry-inequivalent structures may be formed with one Frenkel pair in a starting configuration, one where the carbon atom is displaced into a nearest-neighbor interstital position and the other where it goes to a next-nearest-neighbor position. The latter is shown in Figure \ref{fig2}(b). In both cases, the Frenkel C pair formation near a Ti vacancy in TiC is energetically favorable: the total energy is lowered, respectively, by 1.5 eV or 1.9 eV. The energy gain is due to the strong C--C bonds between the displaced C atom and its C nearest neighbors. Structure 1A is nearly flat, with 3 equally long bonds of 149 pm and 3 bond angles of about 120$^{\circ}$ each. This reconstructed vacancy configuration produces characteristic defect states in the forbidden energy regions of TiC, one below the LVB and the other in the gap between the LVB and UVB. 

\paragraph*{Structure 2G} Shown in Fig. \ref{fig2}(c) is the ground state structure of a Ti vacancy in TiC; it results from two initial configurations with two Frenkel C pairs as well as from one initial configuration with tree displaced C atoms (one of which goes back to a regular lattice site during the relaxation). The total energy of 2G structure is 3.5 eV lower than that of unreconstructed structure 0. The superb stability of this planar "graphene-like" structure inside a Ti vacancy is due to its five C--C bonds (one 151 pm and four 149 pm long; the bond angle between two shorter bonds is 127$^{\circ}$). The electronic structure of 2G has characteristic defect states in the forbidden energy regions: two states below the LVB; a degenerate state between the LVB and UVB. 

The lowering of vacancy formation energy by 3.5 eV is enough to make the usual vacancy mechanism of metal diffusion in TiC competitive with the cluster mechanisms mentioned above and explored in previous studies \citep{Razumov13,Wsun19}. One obstacle is that the vacancy site is blocked by a very strong carbon-bonded structure. Therefore, it is necessary to extend the search further to identify low-energy metastable structures (beyond structure 1A) allowing for metal atom--vacancy exchanges.


\paragraph*{Structure 2I} This asymmetric structure corresponds to two Frenkel C pairs inside a Ti vacancy. The energy of this metastable structure is lower by 2.0 eV than the energy of unreconstructed vacancy structure 0. The two displaced C atoms form five C--C bonds with the nearby C atoms (tree bonds are 146 pm, one 142 pm, and one 152 pm). The bond angles range from 91$^{\circ}$ to 130$^{\circ}$. The the structure is the lowest of several similar (in energy and geometry) asymmetric carbon-bonded structures that are displaced away from the vacancy center, thus leaving the vacant site open for an atomic jump, see Fig. \ref{fig2}(d). The electronic structure of 2I has two characteristic peaks below LVB and four in the gap between LVB and UVB.

\paragraph*{Structure 3C} Even 3 Frenkel C pairs near a metal vacancy can produce metastable structures with a negative energy of reconstruction. For example, the energy of structure 3C shown in Figure \ref{fig2}(e) is some 0.5 eV below zero. The structure is an open chain of seven carbon atoms, four near their regular lattice sites and three in interstital positions. The lengths of four inner C--C bonds are close to 146 pm; the two bonds at the ends have lengths of 142 and 155 pm, respectively. The bond angles range from 89 to 123$^{\circ}$, with an average of 110$^{\circ}$. As for the reconstructed vacancy structures described above, C--C bonds create multiple defect states in the forbidden energy regions of TiC electronic spectrum.

\begin{figure}
\includegraphics[width=9cm]{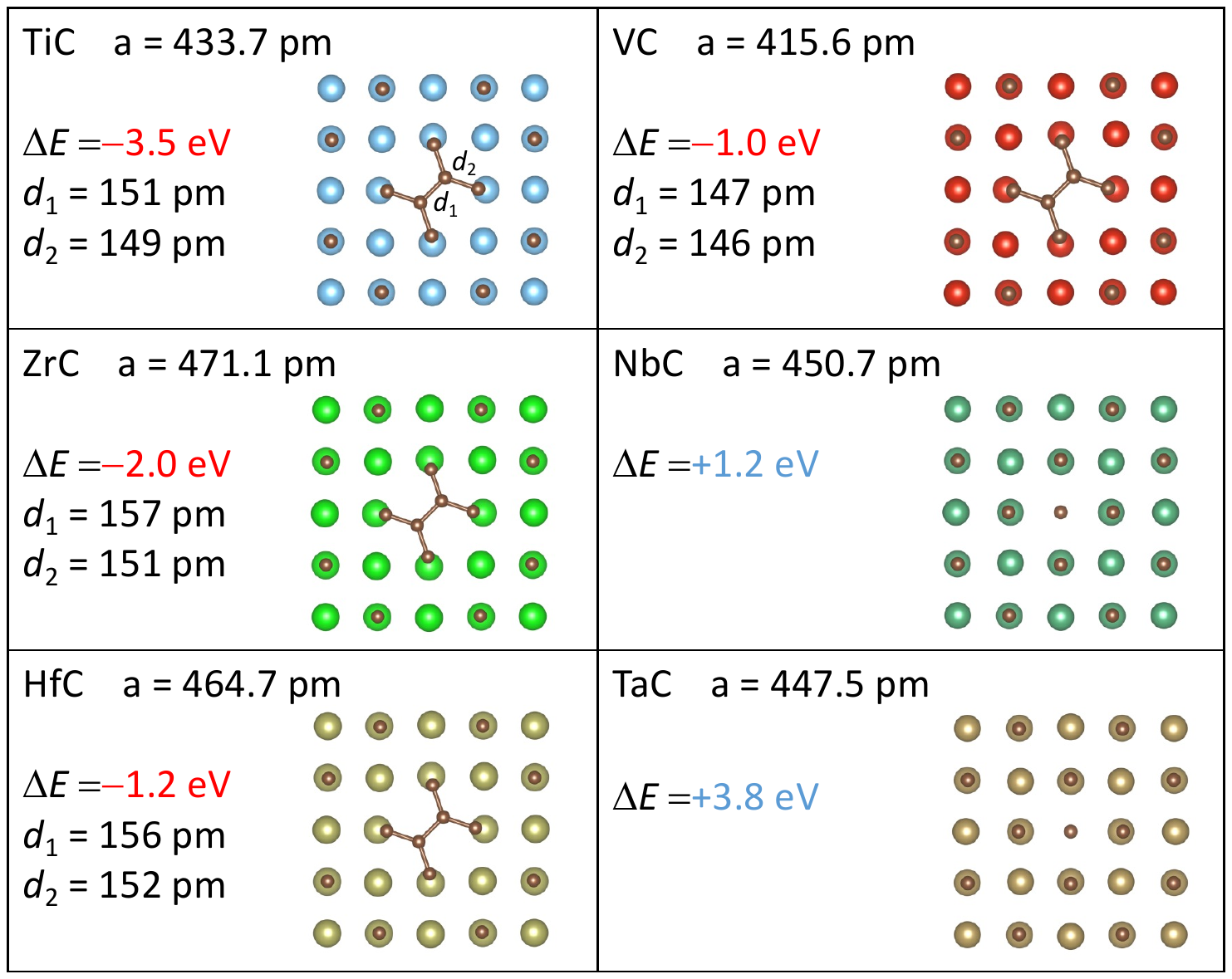}%
\caption{Top views, reconstruction energy $\Delta E = E_{\rm 2G} - E_0$, and C--C bond lengths for the ground state structures of metal vacancies in group 4 and 5 TM carbides: Structure 2G for TiC, ZrC, HfC, and VC; structure 0 for NbC and TaC. Larger spheres denote metal atoms.  \label{fig3}}
\end{figure}

Most structures obtained by relaxing the starting configurations with four or more Frenkel C pairs around a Ti vacancy have energies above zero and, therefore, will not be considered here. Let us now focus our attention on the ground-state structure 2G and the unreconstructed structure 0 to consider their relative stability in cubic carbides of other transition metals of group 4 (Zr, Hf) and 5 (V, Nb, Ta). Figure \ref{fig3} summarizes the atomic geometry of metal vacancies in the considered cubic carbides. 

The planar 2G structure, with two C atoms displaced into the vacancy to form a dimer and C--C bonds with four C atoms in the regular lattice sites, is the ground state of metal vacancies in all group 4 carbides (TiC, ZrC, and HfC). Within this group, the stability of 2G structure decreases with increasing the TM's atomic number. A similar trend is exhibited by metal vacancies in group 5 carbides, where a metal vacancy in VC is calculated to have the reconstructed 2G geometry, while the ground state of metal vacancies in NbC and HfC is the unreconstructed structure 0. The calculated trends show that structural stability of metal vacancies in cubic carbides is influenced by the atomic size factor as well as by the electronic factor (the number of valence electrons): Increasing the atomic size or the valence number of TM atoms destabilizes the carbon-bonded 2G structure inside a metal vacancy in group 4 and 5 TM carbides.

Carbon-bonded structures have been reported to form inside vacancies in Fe--C solid solutions \cite{Clemens2006,Paxton2013} and on surfaces of carbon-enriched nanoclusters (met-cars)\cite{Guo1992,Sofo2012}. Our present calculations show that such structures are more abundant in the solid state and should occur, for example, in some cubic carbides containing metal vacancies. The structural degrees of freedom inside a Ti vacancy have been thoroughly explored using molecular statics to identify a multitude of metastable structures stabilized by the C--C bonds. The planar 2G structure, with a carbon dimer inside the vacancy, is 3.5 eV lower in energy than the usual unreconstructed vacancy geometry. 

Our calculations show an energy gap of 1.5 eV between the ground-state structure and several other C--C bonded structures that leave the vacancy site open for an incoming metal atom. Atomic dynamics at metal vacancies, such as the C--C dimer rotation, may be activated thermally at a thermal energy much smaller than the structural energy gap. These dynamical degrees of freedom deserve a separate consideration as they are crucial for understanding the mechanism of metal diffusion in cubic carbides. 

P.K. acknowledges fruitful discussions with Rolf Sandstr\"om, Tony Paxton, and Alex Zunger. This work has been performed at the Vinnova Competence Centre "Hero-m 2i" financed by the Swedish Governmental Agency for Innovation Systems (Vinnova, grant 2016-00668), Swedish industry, and KTH Royal Institute of Technology. M.N. thanks financial support from the Foundation for Applied Thermodynamics, Sweden. The main body of computations were performed on resources provided by the National Academic Infrastructure for Supercomputing in Sweden (NAISS) and by the Swedish National Infrastructure for Computing (SNIC) at the National Supercomputer Center (NSC) in Link{\"o}ping and at the Center for High Performance Computing (PDC) in Stockholm, partially funded by the Swedish Research Council through grant agreements  no. 2022-06725 and no. 2018-05973. M.N. acknowledges access to computational resources provided by the Hillert Modeling Laboratory funded by the Hugo Carlssons Stiftelse f{\"o}r vetenskaplig forskning, Sweden.
\newline


\nocite{*}

\bibliography{MeC_mono}

\providecommand{\noopsort}[1]{}\providecommand{\singleletter}[1]{#1}%
\begin{thebibliography}{49}%
\makeatletter
\providecommand \@ifxundefined [1]{%
 \@ifx{#1\undefined}
}%
\providecommand \@ifnum [1]{%
 \ifnum #1\expandafter \@firstoftwo
 \else \expandafter \@secondoftwo
 \fi
}%
\providecommand \@ifx [1]{%
 \ifx #1\expandafter \@firstoftwo
 \else \expandafter \@secondoftwo
 \fi
}%
\providecommand \natexlab [1]{#1}%
\providecommand \enquote  [1]{``#1''}%
\providecommand \bibnamefont  [1]{#1}%
\providecommand \bibfnamefont [1]{#1}%
\providecommand \citenamefont [1]{#1}%
\providecommand \href@noop [0]{\@secondoftwo}%
\providecommand \href [0]{\begingroup \@sanitize@url \@href}%
\providecommand \@href[1]{\@@startlink{#1}\@@href}%
\providecommand \@@href[1]{\endgroup#1\@@endlink}%
\providecommand \@sanitize@url [0]{\catcode `\\12\catcode `\$12\catcode
  `\&12\catcode `\#12\catcode `\^12\catcode `\_12\catcode `\%12\relax}%
\providecommand \@@startlink[1]{}%
\providecommand \@@endlink[0]{}%
\providecommand \url  [0]{\begingroup\@sanitize@url \@url }%
\providecommand \@url [1]{\endgroup\@href {#1}{\urlprefix }}%
\providecommand \urlprefix  [0]{URL }%
\providecommand \Eprint [0]{\href }%
\providecommand \doibase [0]{https://doi.org/}%
\providecommand \selectlanguage [0]{\@gobble}%
\providecommand \bibinfo  [0]{\@secondoftwo}%
\providecommand \bibfield  [0]{\@secondoftwo}%
\providecommand \translation [1]{[#1]}%
\providecommand \BibitemOpen [0]{}%
\providecommand \bibitemStop [0]{}%
\providecommand \bibitemNoStop [0]{.\EOS\space}%
\providecommand \EOS [0]{\spacefactor3000\relax}%
\providecommand \BibitemShut  [1]{\csname bibitem#1\endcsname}%
\let\auto@bib@innerbib\@empty
\bibitem [{\citenamefont {Toth}(1971)}]{Toth71}%
  \BibitemOpen
  \bibfield  {author} {\bibinfo {author} {\bibfnamefont {L.~E.}\ \bibnamefont
  {Toth}},\ }\href@noop {} {\emph {\bibinfo {title} {Transition Metal Carbides
  and Nitrides}}}\ (\bibinfo  {publisher} {Academic Press, New York},\ \bibinfo
  {year} {1971})\BibitemShut {NoStop}%
\bibitem [{\citenamefont {Shabalin}(2019)}]{Shabalin2019}%
  \BibitemOpen
  \bibfield  {author} {\bibinfo {author} {\bibfnamefont {I.~L.}\ \bibnamefont
  {Shabalin}},\ }\href@noop {} {\emph {\bibinfo {title} {Ultra-High Temperature
  Materials II: Refractory Carbides I (Ta, Hf, Nb and Zr Carbides)}}}\
  (\bibinfo  {publisher} {Springer Nature, Singapore},\ \bibinfo {year}
  {2019})\BibitemShut {NoStop}%
\bibitem [{\citenamefont {Shabalin}(2020)}]{Shabalin2020}%
  \BibitemOpen
  \bibfield  {author} {\bibinfo {author} {\bibfnamefont {I.~L.}\ \bibnamefont
  {Shabalin}},\ }\href@noop {} {\emph {\bibinfo {title} {Ultra-High Temperature
  Materials III: Refractory Carbides II (Ti and V Carbides)}}}\ (\bibinfo
  {publisher} {Springer Nature, Dordrecht},\ \bibinfo {year}
  {2020})\BibitemShut {NoStop}%
\bibitem [{\citenamefont {Holleck}(1986)}]{Holleck86}%
  \BibitemOpen
  \bibfield  {author} {\bibinfo {author} {\bibfnamefont {H.}~\bibnamefont
  {Holleck}},\ }\href@noop {} {\bibfield  {journal} {\bibinfo  {journal} {J.
  Vac. Sci. Technol. A}\ }\textbf {\bibinfo {volume} {4}},\ \bibinfo {pages}
  {2661} (\bibinfo {year} {1986})}\BibitemShut {NoStop}%
\bibitem [{\citenamefont {Williams}(1997)}]{WsWill97}%
  \BibitemOpen
  \bibfield  {author} {\bibinfo {author} {\bibfnamefont {W.~S.}\ \bibnamefont
  {Williams}},\ }\href@noop {} {\bibfield  {journal} {\bibinfo  {journal}
  {JOM}\ }\textbf {\bibinfo {volume} {49}},\ \bibinfo {pages} {38} (\bibinfo
  {year} {1997})}\BibitemShut {NoStop}%
\bibitem [{\citenamefont {Fahrenholtz}\ \emph {et~al.}(2014)\citenamefont
  {Fahrenholtz}, \citenamefont {Wuchina}, \citenamefont {Lee},\ and\
  \citenamefont {Zhou}}]{Fahrenholtz2014}%
  \BibitemOpen
  \bibinfo {editor} {\bibfnamefont {W.~G.}\ \bibnamefont {Fahrenholtz}},
  \bibinfo {editor} {\bibfnamefont {E.~J.}\ \bibnamefont {Wuchina}}, \bibinfo
  {editor} {\bibfnamefont {W.~E.}\ \bibnamefont {Lee}},\ and\ \bibinfo {editor}
  {\bibfnamefont {Y.}~\bibnamefont {Zhou}},\ eds.,\ \href@noop {} {\emph
  {\bibinfo {title} {Ultra-High Temperature Ceramics: Materials for Extreme
  Environment Applications}}}\ (\bibinfo  {publisher} {Wiley, New Jersey},\
  \bibinfo {year} {2014})\BibitemShut {NoStop}%
\bibitem [{\citenamefont {Keihn}\ and\ \citenamefont
  {Kebler}(1964)}]{KKebler64}%
  \BibitemOpen
  \bibfield  {author} {\bibinfo {author} {\bibfnamefont {F.}~\bibnamefont
  {Keihn}}\ and\ \bibinfo {author} {\bibfnamefont {R.}~\bibnamefont {Kebler}},\
  }\href@noop {} {\bibfield  {journal} {\bibinfo  {journal} {J. Less-common
  Met.}\ }\textbf {\bibinfo {volume} {6}},\ \bibinfo {pages} {484} (\bibinfo
  {year} {1964})}\BibitemShut {NoStop}%
\bibitem [{\citenamefont {Spivak}\ \emph {et~al.}(1974)\citenamefont {Spivak},
  \citenamefont {Andrievskii}, \citenamefont {Rystsov},\ and\ \citenamefont
  {Klimenko}}]{ISpivak74}%
  \BibitemOpen
  \bibfield  {author} {\bibinfo {author} {\bibfnamefont {I.}~\bibnamefont
  {Spivak}}, \bibinfo {author} {\bibfnamefont {R.}~\bibnamefont {Andrievskii}},
  \bibinfo {author} {\bibfnamefont {V.}~\bibnamefont {Rystsov}},\ and\ \bibinfo
  {author} {\bibfnamefont {V.}~\bibnamefont {Klimenko}},\ }\href@noop {}
  {\bibfield  {journal} {\bibinfo  {journal} {Powder Metall. Met. Ceram.}\
  }\textbf {\bibinfo {volume} {13}},\ \bibinfo {pages} {574} (\bibinfo {year}
  {1974})}\BibitemShut {NoStop}%
\bibitem [{\citenamefont {Chermant}\ \emph {et~al.}(1980)\citenamefont
  {Chermant}, \citenamefont {Leclerc},\ and\ \citenamefont
  {Mordike}}]{JChermant80}%
  \BibitemOpen
  \bibfield  {author} {\bibinfo {author} {\bibfnamefont {J.}~\bibnamefont
  {Chermant}}, \bibinfo {author} {\bibfnamefont {G.}~\bibnamefont {Leclerc}},\
  and\ \bibinfo {author} {\bibfnamefont {B.~L.}\ \bibnamefont {Mordike}},\
  }\href@noop {} {\bibfield  {journal} {\bibinfo  {journal} {Z. Metallkde}\
  }\textbf {\bibinfo {volume} {71}},\ \bibinfo {pages} {465} (\bibinfo {year}
  {1980})}\BibitemShut {NoStop}%
\bibitem [{\citenamefont {Smith}\ \emph {et~al.}(2018)\citenamefont {Smith},
  \citenamefont {Ross}, \citenamefont {Leon}, \citenamefont {Weinberger},\ and\
  \citenamefont {Thompson}}]{CJSmith18}%
  \BibitemOpen
  \bibfield  {author} {\bibinfo {author} {\bibfnamefont {C.~J.}\ \bibnamefont
  {Smith}}, \bibinfo {author} {\bibfnamefont {M.~A.}\ \bibnamefont {Ross}},
  \bibinfo {author} {\bibfnamefont {N.~D.}\ \bibnamefont {Leon}}, \bibinfo
  {author} {\bibfnamefont {C.~R.}\ \bibnamefont {Weinberger}},\ and\ \bibinfo
  {author} {\bibfnamefont {G.~B.}\ \bibnamefont {Thompson}},\ }\href@noop {}
  {\bibfield  {journal} {\bibinfo  {journal} {J. Eur. Ceram. Soc.}\ }\textbf
  {\bibinfo {volume} {38}},\ \bibinfo {pages} {5319} (\bibinfo {year}
  {2018})}\BibitemShut {NoStop}%
\bibitem [{\citenamefont {Knotek}\ and\ \citenamefont
  {Barimani}(1989)}]{OKnotek89}%
  \BibitemOpen
  \bibfield  {author} {\bibinfo {author} {\bibfnamefont {O.}~\bibnamefont
  {Knotek}}\ and\ \bibinfo {author} {\bibfnamefont {A.}~\bibnamefont
  {Barimani}},\ }\href@noop {} {\bibfield  {journal} {\bibinfo  {journal} {Thin
  Solid Films}\ }\textbf {\bibinfo {volume} {174}},\ \bibinfo {pages} {51}
  (\bibinfo {year} {1989})}\BibitemShut {NoStop}%
\bibitem [{\citenamefont {Ma}\ \emph {et~al.}(2016)\citenamefont {Ma},
  \citenamefont {Borrajo-Pelaez}, \citenamefont {Hedstr{\"o}m}, \citenamefont
  {Borgh}, \citenamefont {Blomqvist}, \citenamefont {Norgren},\ and\
  \citenamefont {Odqvist}}]{Ma2016}%
  \BibitemOpen
  \bibfield  {author} {\bibinfo {author} {\bibfnamefont {T.}~\bibnamefont
  {Ma}}, \bibinfo {author} {\bibfnamefont {R.}~\bibnamefont {Borrajo-Pelaez}},
  \bibinfo {author} {\bibfnamefont {P.}~\bibnamefont {Hedstr{\"o}m}}, \bibinfo
  {author} {\bibfnamefont {I.}~\bibnamefont {Borgh}}, \bibinfo {author}
  {\bibfnamefont {A.}~\bibnamefont {Blomqvist}}, \bibinfo {author}
  {\bibfnamefont {S.}~\bibnamefont {Norgren}},\ and\ \bibinfo {author}
  {\bibfnamefont {J.}~\bibnamefont {Odqvist}},\ }\href@noop {} {\bibfield
  {journal} {\bibinfo  {journal} {Int. J. Refract. Hard Met.}\ }\textbf
  {\bibinfo {volume} {61}},\ \bibinfo {pages} {238} (\bibinfo {year}
  {2016})}\BibitemShut {NoStop}%
\bibitem [{\citenamefont {Yildiz}\ \emph {et~al.}(2022)\citenamefont {Yildiz},
  \citenamefont {Yixuan}, \citenamefont {Babu}, \citenamefont {Hansen},
  \citenamefont {Eriksson}, \citenamefont {Reddy},\ and\ \citenamefont
  {Hedstr{\"o}m}}]{Yildiz2022}%
  \BibitemOpen
  \bibfield  {author} {\bibinfo {author} {\bibfnamefont {A.~B.}\ \bibnamefont
  {Yildiz}}, \bibinfo {author} {\bibfnamefont {H.}~\bibnamefont {Yixuan}},
  \bibinfo {author} {\bibfnamefont {P.}~\bibnamefont {Babu}}, \bibinfo {author}
  {\bibfnamefont {T.~C.}\ \bibnamefont {Hansen}}, \bibinfo {author}
  {\bibfnamefont {M.}~\bibnamefont {Eriksson}}, \bibinfo {author}
  {\bibfnamefont {K.~M.}\ \bibnamefont {Reddy}},\ and\ \bibinfo {author}
  {\bibfnamefont {P.}~\bibnamefont {Hedstr{\"o}m}},\ }\href@noop {} {\bibfield
  {journal} {\bibinfo  {journal} {J. Eur. Ceram. Soc.}\ }\textbf {\bibinfo
  {volume} {42}},\ \bibinfo {pages} {4429} (\bibinfo {year}
  {2022})}\BibitemShut {NoStop}%
\bibitem [{\citenamefont {Sarian}(1968)}]{Ssarian68}%
  \BibitemOpen
  \bibfield  {author} {\bibinfo {author} {\bibfnamefont {S.}~\bibnamefont
  {Sarian}},\ }\href@noop {} {\bibfield  {journal} {\bibinfo  {journal} {J.
  Appl. Phys}\ }\textbf {\bibinfo {volume} {39}},\ \bibinfo {pages} {3305}
  (\bibinfo {year} {1968})}\BibitemShut {NoStop}%
\bibitem [{\citenamefont {Sarian}(1969)}]{Ssarian69}%
  \BibitemOpen
  \bibfield  {author} {\bibinfo {author} {\bibfnamefont {S.}~\bibnamefont
  {Sarian}},\ }\href@noop {} {\bibfield  {journal} {\bibinfo  {journal} {J.
  Appl. Phys}\ }\textbf {\bibinfo {volume} {40}},\ \bibinfo {pages} {3515}
  (\bibinfo {year} {1969})}\BibitemShut {NoStop}%
\bibitem [{\citenamefont {Zagryazkin}(1969)}]{VNZagryazkin69}%
  \BibitemOpen
  \bibfield  {author} {\bibinfo {author} {\bibfnamefont {V.~N.}\ \bibnamefont
  {Zagryazkin}},\ }\href@noop {} {\bibfield  {journal} {\bibinfo  {journal}
  {Phys. Met. Metallogr.}\ }\textbf {\bibinfo {volume} {28}},\ \bibinfo {pages}
  {292} (\bibinfo {year} {1969})}\BibitemShut {NoStop}%
\bibitem [{\citenamefont {Kohlstedt}\ \emph {et~al.}(1970)\citenamefont
  {Kohlstedt}, \citenamefont {Williams},\ and\ \citenamefont
  {Woodhouse}}]{DKohlstedt70}%
  \BibitemOpen
  \bibfield  {author} {\bibinfo {author} {\bibfnamefont {D.}~\bibnamefont
  {Kohlstedt}}, \bibinfo {author} {\bibfnamefont {W.~S.}\ \bibnamefont
  {Williams}},\ and\ \bibinfo {author} {\bibfnamefont {J.~B.}\ \bibnamefont
  {Woodhouse}},\ }\href@noop {} {\bibfield  {journal} {\bibinfo  {journal} {J.
  Appl. Phys.}\ }\textbf {\bibinfo {volume} {41}},\ \bibinfo {pages} {4475}
  (\bibinfo {year} {1970})}\BibitemShut {NoStop}%
\bibitem [{\citenamefont {Andrievskii}\ \emph {et~al.}(1971)\citenamefont
  {Andrievskii}, \citenamefont {Khormov},\ and\ \citenamefont
  {Alekseeva}}]{RAAndriev71}%
  \BibitemOpen
  \bibfield  {author} {\bibinfo {author} {\bibfnamefont {R.~A.}\ \bibnamefont
  {Andrievskii}}, \bibinfo {author} {\bibfnamefont {Y.~F.}\ \bibnamefont
  {Khormov}},\ and\ \bibinfo {author} {\bibfnamefont {I.~S.}\ \bibnamefont
  {Alekseeva}},\ }\href@noop {} {\bibfield  {journal} {\bibinfo  {journal}
  {Fiz. Metal. Metalloved.}\ }\textbf {\bibinfo {volume} {32}},\ \bibinfo
  {pages} {664} (\bibinfo {year} {1971})}\BibitemShut {NoStop}%
\bibitem [{\citenamefont {Loo}\ \emph {et~al.}(1989)\citenamefont {Loo},
  \citenamefont {Wakelkamp}, \citenamefont {Bastin},\ and\ \citenamefont
  {Metselaar}}]{FJJVLoo89}%
  \BibitemOpen
  \bibfield  {author} {\bibinfo {author} {\bibfnamefont {F.~J. J.~V.}\
  \bibnamefont {Loo}}, \bibinfo {author} {\bibfnamefont {W.}~\bibnamefont
  {Wakelkamp}}, \bibinfo {author} {\bibfnamefont {G.~F.}\ \bibnamefont
  {Bastin}},\ and\ \bibinfo {author} {\bibfnamefont {R.}~\bibnamefont
  {Metselaar}},\ }\href@noop {} {\bibfield  {journal} {\bibinfo  {journal}
  {Solid State Ionics}\ }\textbf {\bibinfo {volume} {32}},\ \bibinfo {pages}
  {824} (\bibinfo {year} {1989})}\BibitemShut {NoStop}%
\bibitem [{\citenamefont {Andrievskii}(2011)}]{Andrievskii2011}%
  \BibitemOpen
  \bibfield  {author} {\bibinfo {author} {\bibfnamefont {R.~A.}\ \bibnamefont
  {Andrievskii}},\ }\href@noop {} {\bibfield  {journal} {\bibinfo  {journal}
  {Powder Metall. Met. C+}\ }\textbf {\bibinfo {volume} {50}},\ \bibinfo
  {pages} {2} (\bibinfo {year} {2011})}\BibitemShut {NoStop}%
\bibitem [{\citenamefont {Hultman}(2000)}]{Hultman2000}%
  \BibitemOpen
  \bibfield  {author} {\bibinfo {author} {\bibfnamefont {L.}~\bibnamefont
  {Hultman}},\ }\href@noop {} {\bibfield  {journal} {\bibinfo  {journal}
  {Vacuum}\ }\textbf {\bibinfo {volume} {57}},\ \bibinfo {pages} {1} (\bibinfo
  {year} {2000})}\BibitemShut {NoStop}%
\bibitem [{\citenamefont {Gusev}\ \emph {et~al.}(2001)\citenamefont {Gusev},
  \citenamefont {Rempel},\ and\ \citenamefont {Magerl}}]{Gusev2001}%
  \BibitemOpen
  \bibfield  {author} {\bibinfo {author} {\bibfnamefont {A.~I.}\ \bibnamefont
  {Gusev}}, \bibinfo {author} {\bibfnamefont {A.~A.}\ \bibnamefont {Rempel}},\
  and\ \bibinfo {author} {\bibfnamefont {A.~A.}\ \bibnamefont {Magerl}},\
  }\href@noop {} {\emph {\bibinfo {title} {Disorder and Order in Strongly
  Nonstoichiometric Compounds: Transition Metal Carbides, Nitrides and
  Oxides}}}\ (\bibinfo  {publisher} {Springer, Berlin},\ \bibinfo {year}
  {2001})\BibitemShut {NoStop}%
\bibitem [{\citenamefont {Andersson}\ \emph {et~al.}(2008)\citenamefont
  {Andersson}, \citenamefont {Korzhavyi},\ and\ \citenamefont
  {Johansson}}]{Andersson08}%
  \BibitemOpen
  \bibfield  {author} {\bibinfo {author} {\bibfnamefont {D.~A.}\ \bibnamefont
  {Andersson}}, \bibinfo {author} {\bibfnamefont {P.~A.}\ \bibnamefont
  {Korzhavyi}},\ and\ \bibinfo {author} {\bibfnamefont {B.}~\bibnamefont
  {Johansson}},\ }\href@noop {} {\bibfield  {journal} {\bibinfo  {journal}
  {CALPHAD}\ }\textbf {\bibinfo {volume} {32}},\ \bibinfo {pages} {543}
  (\bibinfo {year} {2008})}\BibitemShut {NoStop}%
\bibitem [{\citenamefont {Watanabe}\ \emph {et~al.}(1967)\citenamefont
  {Watanabe}, \citenamefont {Castles}, \citenamefont {Jostsons},\ and\
  \citenamefont {Malin}}]{Watanabe67}%
  \BibitemOpen
  \bibfield  {author} {\bibinfo {author} {\bibfnamefont {D.}~\bibnamefont
  {Watanabe}}, \bibinfo {author} {\bibfnamefont {J.}~\bibnamefont {Castles}},
  \bibinfo {author} {\bibfnamefont {A.}~\bibnamefont {Jostsons}},\ and\
  \bibinfo {author} {\bibfnamefont {A.}~\bibnamefont {Malin}},\ }\href@noop {}
  {\bibfield  {journal} {\bibinfo  {journal} {Acta Cryst.}\ }\textbf {\bibinfo
  {volume} {23}},\ \bibinfo {pages} {307} (\bibinfo {year} {1967})}\BibitemShut
  {NoStop}%
\bibitem [{\citenamefont {Valeeva}\ \emph {et~al.}(2000)\citenamefont
  {Valeeva}, \citenamefont {Rempel’},\ and\ \citenamefont
  {Gusev}}]{Valeeva2000}%
  \BibitemOpen
  \bibfield  {author} {\bibinfo {author} {\bibfnamefont {A.~A.}\ \bibnamefont
  {Valeeva}}, \bibinfo {author} {\bibfnamefont {A.~A.}\ \bibnamefont
  {Rempel’}},\ and\ \bibinfo {author} {\bibfnamefont {A.~I.}\ \bibnamefont
  {Gusev}},\ }\href@noop {} {\bibfield  {journal} {\bibinfo  {journal} {JETP
  Letters}\ }\textbf {\bibinfo {volume} {71}},\ \bibinfo {pages} {460}
  (\bibinfo {year} {2000})}\BibitemShut {NoStop}%
\bibitem [{\citenamefont {Valeeva}\ \emph {et~al.}(2001)\citenamefont
  {Valeeva}, \citenamefont {Rempel’},\ and\ \citenamefont
  {Gusev}}]{Valeeva2001}%
  \BibitemOpen
  \bibfield  {author} {\bibinfo {author} {\bibfnamefont {A.~A.}\ \bibnamefont
  {Valeeva}}, \bibinfo {author} {\bibfnamefont {A.~A.}\ \bibnamefont
  {Rempel’}},\ and\ \bibinfo {author} {\bibfnamefont {A.~I.}\ \bibnamefont
  {Gusev}},\ }\href@noop {} {\bibfield  {journal} {\bibinfo  {journal} {Inorg.
  Mater.}\ }\textbf {\bibinfo {volume} {37}},\ \bibinfo {pages} {603} (\bibinfo
  {year} {2001})}\BibitemShut {NoStop}%
\bibitem [{\citenamefont {Andersson}\ \emph {et~al.}(2005)\citenamefont
  {Andersson}, \citenamefont {Korzhavyi},\ and\ \citenamefont
  {Johansson}}]{Andersson05}%
  \BibitemOpen
  \bibfield  {author} {\bibinfo {author} {\bibfnamefont {D.~A.}\ \bibnamefont
  {Andersson}}, \bibinfo {author} {\bibfnamefont {P.~A.}\ \bibnamefont
  {Korzhavyi}},\ and\ \bibinfo {author} {\bibfnamefont {B.}~\bibnamefont
  {Johansson}},\ }\href@noop {} {\bibfield  {journal} {\bibinfo  {journal}
  {Phys. Rev. B}\ }\textbf {\bibinfo {volume} {71}},\ \bibinfo {pages} {144101}
  (\bibinfo {year} {2005})}\BibitemShut {NoStop}%
\bibitem [{\citenamefont {Tsetseris}\ \emph {et~al.}(2011)\citenamefont
  {Tsetseris}, \citenamefont {Logothetidis},\ and\ \citenamefont
  {Pantelides}}]{Tsetseris2010}%
  \BibitemOpen
  \bibfield  {author} {\bibinfo {author} {\bibfnamefont {L.}~\bibnamefont
  {Tsetseris}}, \bibinfo {author} {\bibfnamefont {S.}~\bibnamefont
  {Logothetidis}},\ and\ \bibinfo {author} {\bibfnamefont {S.~T.}\ \bibnamefont
  {Pantelides}},\ }\href@noop {} {\bibfield  {journal} {\bibinfo  {journal}
  {Surf. Coat. Technol.}\ }\textbf {\bibinfo {volume} {204}},\ \bibinfo {pages}
  {2089} (\bibinfo {year} {2011})}\BibitemShut {NoStop}%
\bibitem [{\citenamefont {Razumovskiy}\ \emph {et~al.}(2015)\citenamefont
  {Razumovskiy}, \citenamefont {Popov}, \citenamefont {Ding},\ and\
  \citenamefont {Odqvist}}]{Razumov15}%
  \BibitemOpen
  \bibfield  {author} {\bibinfo {author} {\bibfnamefont {V.~I.}\ \bibnamefont
  {Razumovskiy}}, \bibinfo {author} {\bibfnamefont {M.~N.}\ \bibnamefont
  {Popov}}, \bibinfo {author} {\bibfnamefont {H.}~\bibnamefont {Ding}},\ and\
  \bibinfo {author} {\bibfnamefont {J.}~\bibnamefont {Odqvist}},\ }\href@noop
  {} {\bibfield  {journal} {\bibinfo  {journal} {Comput. Mater. Sci.}\ }\textbf
  {\bibinfo {volume} {104}},\ \bibinfo {pages} {147} (\bibinfo {year}
  {2015})}\BibitemShut {NoStop}%
\bibitem [{\citenamefont {Gambino}\ \emph {et~al.}(2017)\citenamefont
  {Gambino}, \citenamefont {Sangiovanni}, \citenamefont {Alling},\ and\
  \citenamefont {Abrikosov}}]{Gambino2017}%
  \BibitemOpen
  \bibfield  {author} {\bibinfo {author} {\bibfnamefont {D.}~\bibnamefont
  {Gambino}}, \bibinfo {author} {\bibfnamefont {D.~G.}\ \bibnamefont
  {Sangiovanni}}, \bibinfo {author} {\bibfnamefont {B.}~\bibnamefont
  {Alling}},\ and\ \bibinfo {author} {\bibfnamefont {I.~A.}\ \bibnamefont
  {Abrikosov}},\ }\href@noop {} {\bibfield  {journal} {\bibinfo  {journal}
  {Phys. Rev. B}\ }\textbf {\bibinfo {volume} {96}},\ \bibinfo {pages} {104306}
  (\bibinfo {year} {2017})}\BibitemShut {NoStop}%
\bibitem [{\citenamefont {Tsetseris}\ \emph {et~al.}(2008)\citenamefont
  {Tsetseris}, \citenamefont {Logothetidis},\ and\ \citenamefont
  {Pantelides}}]{Ltsets08}%
  \BibitemOpen
  \bibfield  {author} {\bibinfo {author} {\bibfnamefont {L.}~\bibnamefont
  {Tsetseris}}, \bibinfo {author} {\bibfnamefont {S.}~\bibnamefont
  {Logothetidis}},\ and\ \bibinfo {author} {\bibfnamefont {S.~T.}\ \bibnamefont
  {Pantelides}},\ }\href@noop {} {\bibfield  {journal} {\bibinfo  {journal}
  {Acta. Mater.}\ }\textbf {\bibinfo {volume} {56}},\ \bibinfo {pages} {2864}
  (\bibinfo {year} {2008})}\BibitemShut {NoStop}%
\bibitem [{\citenamefont {Pinto}\ \emph {et~al.}(2009)\citenamefont {Pinto},
  \citenamefont {Coutinho}, \citenamefont {Ramos}, \citenamefont {Vaz},\ and\
  \citenamefont {Marques}}]{HmPinto09}%
  \BibitemOpen
  \bibfield  {author} {\bibinfo {author} {\bibfnamefont {H.~M.}\ \bibnamefont
  {Pinto}}, \bibinfo {author} {\bibfnamefont {J.}~\bibnamefont {Coutinho}},
  \bibinfo {author} {\bibfnamefont {M.~M.~D.}\ \bibnamefont {Ramos}}, \bibinfo
  {author} {\bibfnamefont {F.}~\bibnamefont {Vaz}},\ and\ \bibinfo {author}
  {\bibfnamefont {L.}~\bibnamefont {Marques}},\ }\href@noop {} {\bibfield
  {journal} {\bibinfo  {journal} {Mat. Sci. Eng. B}\ }\textbf {\bibinfo
  {volume} {165}},\ \bibinfo {pages} {194} (\bibinfo {year}
  {2009})}\BibitemShut {NoStop}%
\bibitem [{\citenamefont {Razumovskiy}\ \emph {et~al.}(2011)\citenamefont
  {Razumovskiy}, \citenamefont {Korzhavyi},\ and\ \citenamefont
  {Ruban}}]{Razumov11}%
  \BibitemOpen
  \bibfield  {author} {\bibinfo {author} {\bibfnamefont {V.~I.}\ \bibnamefont
  {Razumovskiy}}, \bibinfo {author} {\bibfnamefont {P.~A.}\ \bibnamefont
  {Korzhavyi}},\ and\ \bibinfo {author} {\bibfnamefont {A.~V.}\ \bibnamefont
  {Ruban}},\ }\href@noop {} {\bibfield  {journal} {\bibinfo  {journal} {Solid
  State Phenom.}\ }\textbf {\bibinfo {volume} {174}},\ \bibinfo {pages} {990}
  (\bibinfo {year} {2011})}\BibitemShut {NoStop}%
\bibitem [{\citenamefont {Tang}\ \emph {et~al.}(2020)\citenamefont {Tang},
  \citenamefont {Salehin}, \citenamefont {Thompson},\ and\ \citenamefont
  {Weinberger}}]{Xiao20}%
  \BibitemOpen
  \bibfield  {author} {\bibinfo {author} {\bibfnamefont {X.}~\bibnamefont
  {Tang}}, \bibinfo {author} {\bibfnamefont {R.}~\bibnamefont {Salehin}},
  \bibinfo {author} {\bibfnamefont {G.~B.}\ \bibnamefont {Thompson}},\ and\
  \bibinfo {author} {\bibfnamefont {C.~R.}\ \bibnamefont {Weinberger}},\
  }\href@noop {} {\bibfield  {journal} {\bibinfo  {journal} {Phys. Rev.
  Materials}\ }\textbf {\bibinfo {volume} {4}},\ \bibinfo {pages} {093602}
  (\bibinfo {year} {2020})}\BibitemShut {NoStop}%
\bibitem [{\citenamefont {Salehin}\ \emph {et~al.}(2021)\citenamefont
  {Salehin}, \citenamefont {Tang}, \citenamefont {Thompson},\ and\
  \citenamefont {Weinberger}}]{Rofiq21}%
  \BibitemOpen
  \bibfield  {author} {\bibinfo {author} {\bibfnamefont {R.}~\bibnamefont
  {Salehin}}, \bibinfo {author} {\bibfnamefont {X.}~\bibnamefont {Tang}},
  \bibinfo {author} {\bibfnamefont {G.~B.}\ \bibnamefont {Thompson}},\ and\
  \bibinfo {author} {\bibfnamefont {C.~R.}\ \bibnamefont {Weinberger}},\
  }\href@noop {} {\bibfield  {journal} {\bibinfo  {journal} {Comp. Mater.
  Sci.}\ }\textbf {\bibinfo {volume} {199}},\ \bibinfo {pages} {110713}
  (\bibinfo {year} {2021})}\BibitemShut {NoStop}%
\bibitem [{\citenamefont {Razumovskiy}\ \emph {et~al.}(2013)\citenamefont
  {Razumovskiy}, \citenamefont {Ruban}, \citenamefont {Odqvist},\ and\
  \citenamefont {Korzhavyi}}]{Razumov13}%
  \BibitemOpen
  \bibfield  {author} {\bibinfo {author} {\bibfnamefont {V.~I.}\ \bibnamefont
  {Razumovskiy}}, \bibinfo {author} {\bibfnamefont {A.~V.}\ \bibnamefont
  {Ruban}}, \bibinfo {author} {\bibfnamefont {J.}~\bibnamefont {Odqvist}},\
  and\ \bibinfo {author} {\bibfnamefont {P.~A.}\ \bibnamefont {Korzhavyi}},\
  }\href@noop {} {\bibfield  {journal} {\bibinfo  {journal} {Phys. Rev. B}\
  }\textbf {\bibinfo {volume} {87}},\ \bibinfo {pages} {054203} (\bibinfo
  {year} {2013})}\BibitemShut {NoStop}%
\bibitem [{\citenamefont {Momma}\ and\ \citenamefont
  {Izumi}(2011)}]{Momma2011}%
  \BibitemOpen
  \bibfield  {author} {\bibinfo {author} {\bibfnamefont {K.}~\bibnamefont
  {Momma}}\ and\ \bibinfo {author} {\bibfnamefont {F.}~\bibnamefont {Izumi}},\
  }\href@noop {} {\bibfield  {journal} {\bibinfo  {journal} {J. Appl.
  Crystallogr.}\ }\textbf {\bibinfo {volume} {44}},\ \bibinfo {pages} {1272}
  (\bibinfo {year} {2011})}\BibitemShut {NoStop}%
\bibitem [{\citenamefont {Sun}\ \emph {et~al.}(2015)\citenamefont {Sun},
  \citenamefont {Ehteshami},\ and\ \citenamefont {Korzhavyi}}]{Wsun15}%
  \BibitemOpen
  \bibfield  {author} {\bibinfo {author} {\bibfnamefont {W.}~\bibnamefont
  {Sun}}, \bibinfo {author} {\bibfnamefont {H.}~\bibnamefont {Ehteshami}},\
  and\ \bibinfo {author} {\bibfnamefont {P.}~\bibnamefont {Korzhavyi}},\
  }\href@noop {} {\bibfield  {journal} {\bibinfo  {journal} {Phys. Rev. B}\
  }\textbf {\bibinfo {volume} {91}},\ \bibinfo {pages} {134111} (\bibinfo
  {year} {2015})}\BibitemShut {NoStop}%
\bibitem [{\citenamefont {Sun}\ \emph {et~al.}(2019)\citenamefont {Sun},
  \citenamefont {Ehteshami}, \citenamefont {Kent},\ and\ \citenamefont
  {Korzhavyi}}]{Wsun19}%
  \BibitemOpen
  \bibfield  {author} {\bibinfo {author} {\bibfnamefont {W.}~\bibnamefont
  {Sun}}, \bibinfo {author} {\bibfnamefont {H.}~\bibnamefont {Ehteshami}},
  \bibinfo {author} {\bibfnamefont {P.~R.~C.}\ \bibnamefont {Kent}},\ and\
  \bibinfo {author} {\bibfnamefont {P.}~\bibnamefont {Korzhavyi}},\ }\href@noop
  {} {\bibfield  {journal} {\bibinfo  {journal} {Acta Mater.}\ }\textbf
  {\bibinfo {volume} {165}},\ \bibinfo {pages} {381} (\bibinfo {year}
  {2019})}\BibitemShut {NoStop}%
\bibitem [{\citenamefont {Kresse}\ and\ \citenamefont
  {Furthm{\"u}ller}(1996)}]{Kresse96}%
  \BibitemOpen
  \bibfield  {author} {\bibinfo {author} {\bibfnamefont {G.}~\bibnamefont
  {Kresse}}\ and\ \bibinfo {author} {\bibfnamefont {J.}~\bibnamefont
  {Furthm{\"u}ller}},\ }\href@noop {} {\bibfield  {journal} {\bibinfo
  {journal} {Phys. Rev. B}\ }\textbf {\bibinfo {volume} {54}},\ \bibinfo
  {pages} {11169} (\bibinfo {year} {1996})}\BibitemShut {NoStop}%
\bibitem [{\citenamefont {Bl{\"o}chl}(1994)}]{Blochl94}%
  \BibitemOpen
  \bibfield  {author} {\bibinfo {author} {\bibfnamefont {P.~E.}\ \bibnamefont
  {Bl{\"o}chl}},\ }\href@noop {} {\bibfield  {journal} {\bibinfo  {journal}
  {Phys. Rev. B}\ }\textbf {\bibinfo {volume} {50}},\ \bibinfo {pages} {17953}
  (\bibinfo {year} {1994})}\BibitemShut {NoStop}%
\bibitem [{\citenamefont {Kresse}\ and\ \citenamefont
  {Joubert}(1999)}]{Gkresse99}%
  \BibitemOpen
  \bibfield  {author} {\bibinfo {author} {\bibfnamefont {G.}~\bibnamefont
  {Kresse}}\ and\ \bibinfo {author} {\bibfnamefont {D.}~\bibnamefont
  {Joubert}},\ }\href@noop {} {\bibfield  {journal} {\bibinfo  {journal} {Phys.
  Rev. B}\ }\textbf {\bibinfo {volume} {59}},\ \bibinfo {pages} {1758}
  (\bibinfo {year} {1999})}\BibitemShut {NoStop}%
\bibitem [{\citenamefont {Perdew}\ \emph {et~al.}(1996)\citenamefont {Perdew},
  \citenamefont {Burke},\ and\ \citenamefont {Ernzerhof}}]{Perdew96}%
  \BibitemOpen
  \bibfield  {author} {\bibinfo {author} {\bibfnamefont {J.~P.}\ \bibnamefont
  {Perdew}}, \bibinfo {author} {\bibfnamefont {K.}~\bibnamefont {Burke}},\ and\
  \bibinfo {author} {\bibfnamefont {M.}~\bibnamefont {Ernzerhof}},\ }\href@noop
  {} {\bibfield  {journal} {\bibinfo  {journal} {Phys. Rev. Letters}\ }\textbf
  {\bibinfo {volume} {77}},\ \bibinfo {pages} {3865} (\bibinfo {year}
  {1996})}\BibitemShut {NoStop}%
\bibitem [{\citenamefont {Rempel}\ \emph {et~al.}(1998)\citenamefont {Rempel},
  \citenamefont {Zueva}, \citenamefont {Lipatnikov},\ and\ \citenamefont
  {Schaefer}}]{Rempel1998}%
  \BibitemOpen
  \bibfield  {author} {\bibinfo {author} {\bibfnamefont {A.~A.}\ \bibnamefont
  {Rempel}}, \bibinfo {author} {\bibfnamefont {L.~V.}\ \bibnamefont {Zueva}},
  \bibinfo {author} {\bibfnamefont {V.~N.}\ \bibnamefont {Lipatnikov}},\ and\
  \bibinfo {author} {\bibfnamefont {H.-E.}\ \bibnamefont {Schaefer}},\
  }\href@noop {} {\bibfield  {journal} {\bibinfo  {journal} {Phys. Stat. Sol.
  (a)}\ }\textbf {\bibinfo {volume} {169}},\ \bibinfo {pages} {R9} (\bibinfo
  {year} {1998})}\BibitemShut {NoStop}%
\bibitem [{\citenamefont {Valeeva}\ \emph {et~al.}(2007)\citenamefont
  {Valeeva}, \citenamefont {Rempel}, \citenamefont {Sprengel},\ and\
  \citenamefont {Schaefer}}]{Valeeva2007}%
  \BibitemOpen
  \bibfield  {author} {\bibinfo {author} {\bibfnamefont {A.}~\bibnamefont
  {Valeeva}}, \bibinfo {author} {\bibfnamefont {A.}~\bibnamefont {Rempel}},
  \bibinfo {author} {\bibfnamefont {W.}~\bibnamefont {Sprengel}},\ and\
  \bibinfo {author} {\bibfnamefont {H.-E.}\ \bibnamefont {Schaefer}},\
  }\href@noop {} {\bibfield  {journal} {\bibinfo  {journal} {Phys. Rev. B}\
  }\textbf {\bibinfo {volume} {75}},\ \bibinfo {pages} {094107} (\bibinfo
  {year} {2007})}\BibitemShut {NoStop}%
\bibitem [{\citenamefont {F{\"o}rst}\ \emph {et~al.}(2006)\citenamefont
  {F{\"o}rst}, \citenamefont {Slycke}, \citenamefont {{Van Vliet}},\ and\
  \citenamefont {Yip}}]{Clemens2006}%
  \BibitemOpen
  \bibfield  {author} {\bibinfo {author} {\bibfnamefont {C.~J.}\ \bibnamefont
  {F{\"o}rst}}, \bibinfo {author} {\bibfnamefont {J.}~\bibnamefont {Slycke}},
  \bibinfo {author} {\bibfnamefont {K.~J.}\ \bibnamefont {{Van Vliet}}},\ and\
  \bibinfo {author} {\bibfnamefont {S.}~\bibnamefont {Yip}},\ }\href@noop {}
  {\bibfield  {journal} {\bibinfo  {journal} {Phys. Rev. Letters}\ }\textbf
  {\bibinfo {volume} {96}},\ \bibinfo {pages} {175501} (\bibinfo {year}
  {2006})}\BibitemShut {NoStop}%
\bibitem [{\citenamefont {Paxton}\ and\ \citenamefont
  {Els{\"a}sser}(2013)}]{Paxton2013}%
  \BibitemOpen
  \bibfield  {author} {\bibinfo {author} {\bibfnamefont {A.~T.}\ \bibnamefont
  {Paxton}}\ and\ \bibinfo {author} {\bibfnamefont {C.}~\bibnamefont
  {Els{\"a}sser}},\ }\href@noop {} {\bibfield  {journal} {\bibinfo  {journal}
  {Phys. Rev. B}\ }\textbf {\bibinfo {volume} {87}},\ \bibinfo {pages} {224110}
  (\bibinfo {year} {2013})}\BibitemShut {NoStop}%
\bibitem [{\citenamefont {Guo}\ \emph {et~al.}(1992)\citenamefont {Guo},
  \citenamefont {Kerns},\ and\ \citenamefont {{Castleman, Jr.}}}]{Guo1992}%
  \BibitemOpen
  \bibfield  {author} {\bibinfo {author} {\bibfnamefont {B.~C.}\ \bibnamefont
  {Guo}}, \bibinfo {author} {\bibfnamefont {K.~P.}\ \bibnamefont {Kerns}},\
  and\ \bibinfo {author} {\bibfnamefont {A.~W.}\ \bibnamefont {{Castleman,
  Jr.}}},\ }\href@noop {} {\bibfield  {journal} {\bibinfo  {journal} {Science}\
  }\textbf {\bibinfo {volume} {255}},\ \bibinfo {pages} {1411} (\bibinfo {year}
  {1992})}\BibitemShut {NoStop}%
\bibitem [{\citenamefont {Berkdemir}\ \emph {et~al.}(2012)\citenamefont
  {Berkdemir}, \citenamefont {{Castleman, Jr.}},\ and\ \citenamefont
  {Sofo}}]{Sofo2012}%
  \BibitemOpen
  \bibfield  {author} {\bibinfo {author} {\bibfnamefont {C.}~\bibnamefont
  {Berkdemir}}, \bibinfo {author} {\bibfnamefont {A.~W.}\ \bibnamefont
  {{Castleman, Jr.}}},\ and\ \bibinfo {author} {\bibfnamefont {J.~O.}\
  \bibnamefont {Sofo}},\ }\href@noop {} {\bibfield  {journal} {\bibinfo
  {journal} {Phys. Chem. Chem. Phys.}\ }\textbf {\bibinfo {volume} {14}},\
  \bibinfo {pages} {9642} (\bibinfo {year} {2012})}\BibitemShut {NoStop}%
\end{thebibliography}%

\end{document}